\documentclass[%
 reprint,
superscriptaddress,
 amsmath,amssymb,
 aps,
]{revtex4-2}

\usepackage{graphicx}
\usepackage{dcolumn}
\usepackage{bm}
\usepackage{xcolor}
\usepackage[normalem]{ulem}
\usepackage{lipsum}  
\usepackage{amsmath}

\usepackage{pdfpages} 
 \usepackage{pgffor} 

\makeatletter
\AtBeginDocument{\let\LS@rot\@undefined}
\makeatother

\begin{document}

\newcommand{\ZS}{Z_{\text{s}}}

\preprint{APS/123-QED}

\title{Accelerating field decay along nonlocal metasurfaces by suppressing the Norton wave}

\author{Alexander Zhuravlev}%
 \email{a.zhuravlev@metalab.ifmo.ru}
\affiliation{School of Physics and Engineering, ITMO University, St. Petersburg, Russia}%

\author{Dmitry Tatarnikov}
   \affiliation{School of Physics and Engineering, ITMO University, St. Petersburg, Russia}
   
\author{Yury Kurenkov}%
  \affiliation{School of Physics and Engineering, ITMO University, St. Petersburg, Russia}
  
\author{Stanislav Glybovski}%
  \affiliation{School of Physics and Engineering, ITMO University, St. Petersburg, Russia}



\begin{abstract}
Studying the nature of electromagnetic fields of dipole sources over a homogeneous flat ground or impedance surfaces has a long history. In general, at a long distance $r$ from the source, the near-surface field is mostly contributed by the geometrical optics term (describing the radiation pattern), a guided wave, and the higher-order reactive contribution referred to as the Norton wave. In the special case of a perfect magnetic conductor interface, the first two terms vanish, so the residual Norton wave determines the steepest achievable field decay profile of $~r^{-3/2}$ (for a two-dimensional horizontal magnetic dipole). In this letter, we reveal that in the presence of a nonlocal metasurface described by the second-order impedance boundary condition, the field decay can be further accelerated by suppressing the Norton wave (approaching the profiles $r^{-5/2}$ and $r^{-7/2}$ for electric and magnetic fields, respectively). In a proposed practical realization of a nonlocal metasurface, the effect is numerically verified and shown to reduce the edge diffraction effects by 10 dB for the shield diameter of only one wavelength, paving the way toward compact antenna systems.
\end{abstract}

\maketitle

The electromagnetic field solution for a dipole source above a lossy dielectric boundary (Sommerfeld's problem) is a classical model of an antenna operating over the Earth at radio frequencies \cite{zenneck1907fortpflanzung,sommerfeld1909ausbreitung,norton1936propagation} or a nano-antenna over a plasmonic metal-air interface in the optical regime \cite{lalanne2005theory,Nikitin2009}. The asymptotics of the exact solution (in the form of Sommerfeld-like integrals) commonly used in the literature to analyze the near-surface field behavior at a long distance $r$ from the source, in general, contains the following contributions: (i) the geometrical optics (GO) term (related to the radiation pattern) with a decay profile of $r^{-1}$ (in a 3D problem) or $r^{-1/2}$ (in a 2D problem); (ii) higher-order terms in a power series of $r^{-1}$, including the so-called Norton wave  decaying as $r^{-2}$ (in a 3D problem)~\cite{norton1936propagation} or $r^{-3/2}$ (in a 2D problem)~\cite{Nikitin2009}, which describe reactive fields; and (iii) the guided wave typically having the form of a trapped (surface) wave. The interplay between the above terms substantially defines the near-surface field profile far from the source.

Metasurfaces (MSs; two-dimensional periodic structures with subwavelength periodicity) serve as a platform for synthesizing desired scattered field distributions under the given excitation~\cite{glybovski2016metasurfaces}. 
The local response of impenetrable (impedance) MSs can be described on the macroscopic level with a frequency-dependent surface impedance $Z_{\text{s}}$, which connects the averaged tangential electric $E_\text{t}$ and magnetic $H_\text{t}$ fields at the same point on the MS through the Leontovich boundary condition: $E_{\text{t}}=Z_{\text{s}}H_{\text{t}}$. The most popular realizations of such local MSs are the corrugated \cite{kildal1990artificially} and mushroom-type~\cite{798001} surfaces, for which $|\ZS| \gg \eta$ near the resonance, where $\eta$ is the characteristic impedance of free space. Such so-called high-impedance surfaces (HISs) serve as artificial magnetic conductor (AMC) reflectors for propagating plane waves and simultaneously suppress any propagating surface waves (the electromagnetic band gap (EBG) property) \cite{798001}. 
For a dipole source above a local MS with arbitrary $\ZS$, an asymptotic representation of the near-surface field with the same types of waves as for the lossy dielectric is valid~\cite{wait1957excitation}. 
\begin{figure}
  \centering
  \includegraphics[width=1\columnwidth]{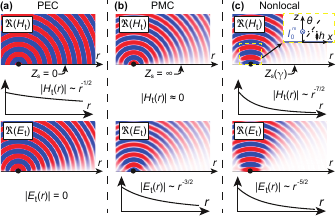} 
  \caption{Fields excited by a HMD placed at height $h\ll\lambda$ above a (a) PEC; (b) PMC and (c) nonlocal impedance boundary with an impedance pole at a grazing angle. The geometry of the 2D problem  considered is depicted in the inset of (c).}
  \label{Fig1:fields}
\end{figure}
One limiting case $\ZS=0$ corresponds to a perfect electric conductor (PEC) ground plane, for which the GO term dominates in the asymptotic behavior of a field excited by an in-plane (located at small height $h$ compared to wavelength $\lambda$) horizontal magnetic dipole (HMD), while no guided wave is supported (see Fig.~\ref{Fig1:fields}(a)). 
In another limit when $\ZS\rightarrow\infty$, mimicking the perfect magnetic conductor (PMC) boundary, the GO term is zero because of the destructive interference between the HMD and its mirror image, while still no propagating guided wave is supported. As a result, the main contribution to the near-surface field is the Norton wave (Fig.~\ref{Fig1:fields}(b)). Due to a faster field decay compared to the PEC ground plane ($r^{-3/2}$ vs. $r^{-1/2}$ for a 2D problem \cite{Tatarnikov2005}), HISs near the resonance serve as compact antenna shields with improved suppression in the shadow domain \cite{Baggen2008,Scire-Scappuzzo2009,Tatarnikov2011} and  decoupling structures \cite{Fan2003}.

For a more general nonlocal MS response, $E_{\text{t}}$ and $H_{\text{t}}$ at different points are related to each other~\cite{Simovski2018}. The nonlocality manifests itself in the spatial dispersion (SD) of $\ZS$, i.e. its variation with the incident wave vector \cite{Luukkonen2008}. Adjusting the law of SD has been shown to control the angular properties of MS absorbers \cite{Zhirihin2017}, antenna reflectors \cite{Zhuravlev2025}, and the transmittance of the anti-reflection coating \cite{zhuravlev2025WAR}. Although the problem of dipole excitation of spatially dispersive wire grids \cite{Lindell1986} and plasmonic metal boundaries \cite{Nikitin2009} was solved analytically, to our knowledge, the possibility of systematically controlling and accelerating the field decay along the MS has not been studied.


To investigate the near-surface field decay in the presence of nonlocality, let us consider a 2D problem with an infinite magnetic line current of complex amplitude $I^{\text{m}}_0$ (a 2D dipole oriented along the $y$ axis) at  height $h\ll \lambda$ above a uniform and isotropic MS. The latter is located in the $XY$ plane and is modeled by a spatially dispersive surface impedance $\ZS$, as shown in the inset of Fig.~\ref{Fig1:fields}(c). The source creates a TM$_y$-polarized field with only $H_y$, $E_x$, and $E_z$ components nonzero.
The tangential components of the total field $E_t=E_x$, $H_t=H_y$ (the sums of the incident and reflected fields), observed at angle $0\le\theta\le90^{\circ}$ and distance $r$ from the source, have the following rigorous spectral representations \cite[Sec.~5]{Felsen1994}:
\begin{equation}    
\label{eq:ift}
    \begin{Bmatrix} E_{x}(\theta) \\ H_{y}(\theta) \end{Bmatrix}
=\frac{C}{\sqrt{2\pi}}\int\limits_{P}f_{\lbrace E_x,H_y\rbrace}(\tau)e^{-j\Omega\cos(\tau-\theta)}d\tau,
\end{equation}
where $\tau$ is related to the normalized tangential wave vector component $\gamma=k_x/k$ of a spatial spectrum harmonic as $\sin\tau=\gamma$ (harmonics with $0
\le\gamma \le 1$ and $0^{\circ}\le\tau=\theta\le90^{\circ}$ are propagating plane waves, while harmonics with $\gamma > 1$ are evanescent plane waves); $C=I^{\text{m}}_0k/\sqrt{2\pi}\eta$; $k$ is the wavenumber of free space; $\Omega=kr$; $r=\sqrt{x^2+(z-h)^2}$; $P$ is the contour of integration (depicted in \cite[Fig.~S.1]{suppmat}); and the integrated functions are determined by the SD law $\ZS(\gamma)$ as:
\begin{equation} \label{eq:ift_func}
    f_{H_y}=\frac{-\cos \tau}{\ZS(\sin\tau)/\eta+\cos \tau}; ~f_{E_x} =-\ZS (\sin\tau)f_{H_y}.
\end{equation}

A far-field approximation ($\Omega\gg2\pi$) of (\ref{eq:ift}) for a specific observation angle $\tau=\theta \le 90^{\circ}$ can be derived using the steepest descent method \cite[Ch. 4]{Felsen1994}:
\begin{widetext}
\begin{equation}
\label{eq:total_ff}
\begin{Bmatrix} E_{x}(\theta) \\ H_{y}(\theta) \end{Bmatrix}
\approx C e^{-j\left(\Omega-\frac{\pi}{4}\right)}\biggl( 
\underbrace{
\begin{Bmatrix} f_{E_x}(\theta) \\ f_{H_y}(\theta) \end{Bmatrix}
\frac{1}{\Omega^{\frac{1}{2}}}
}_{\text{GO term}}
+  \sum_{n=1}^{\infty} 
\underbrace{
j^n f^{(2n)}_{\lbrace E_x,H_y\rbrace}(\theta) 
\frac{(2n-1)!!}{(2n)!}
}_{F_{n,\lbrace E_x,H_y\rbrace}}
\frac{1}{\Omega^{\frac{2n+1}{2}}}\biggr) 
-\underbrace{
\sum_{i}{
\begin{Bmatrix} E_{x,i}^{\text{GW}} \\ H_{y,i}^{\text{GW}} \end{Bmatrix}
e^{-j\Omega\sin{\tau_{i}}}}
}_{\text{Guided waves}},
  \end{equation}
\end{widetext}
where $f^{(2n)}(\theta)$ denotes the derivative of $2n$-th order with respect to $\theta$, while ``GW'' stands for guided waves. 

In (\ref{eq:total_ff}) the first term is the GO term, for which functions (\ref{eq:ift_func}) for $0^{\circ}\le\tau=\theta\le90^{\circ}$ determine the radiation pattern. The next term ($n=1$) is the Norton wave, which together with other higher-order terms ($n\ge2$) exhibits algebraic dependence on the distance $\Omega$. The last sum in (\ref{eq:total_ff}) of terms  exponentially dependent on the distance, resulting from the poles of (\ref{eq:ift_func}) with respect to $\tau$, accounts for the contribution of guided waves with excitation amplitudes $E_{x,i}^{\text{GW}}$, $H_{y,i}^{\text{GW}}$. Depending on the roots $\tau_{i}$ of the characteristic equation, that is, the denominator of (\ref{eq:ift_func}) equals zero, guided waves may be surface waves (SWs for $\sin\tau_i \in \mathbb{R}$) or leaky waves (LWs for $\sin\tau_i \in \mathbb{C}$).

Consider the fields near the MS plane. Both $f_{E_x}$ and $f_{H_y}$ vanish at $\theta\rightarrow90^{\circ}$ for all SD laws except those that satisfy condition $\ZS(90^{\circ})=0$, so that the GO term in (\ref{eq:total_ff}) becomes zero. However, the contribution of other asymptotic terms strongly depends on the SD of $\ZS$. Let us find the specific conditions under which the fields $E_{x}(90^{\circ}),~H_{y}(90^{\circ})$ decay the fastest with $\Omega$. Obviously, no SWs should be supported by the MS, while the Norton wave and LWs should be minimized or canceled. 

\begin{figure*}[t]
  \centering
  \includegraphics[width=1\textwidth]{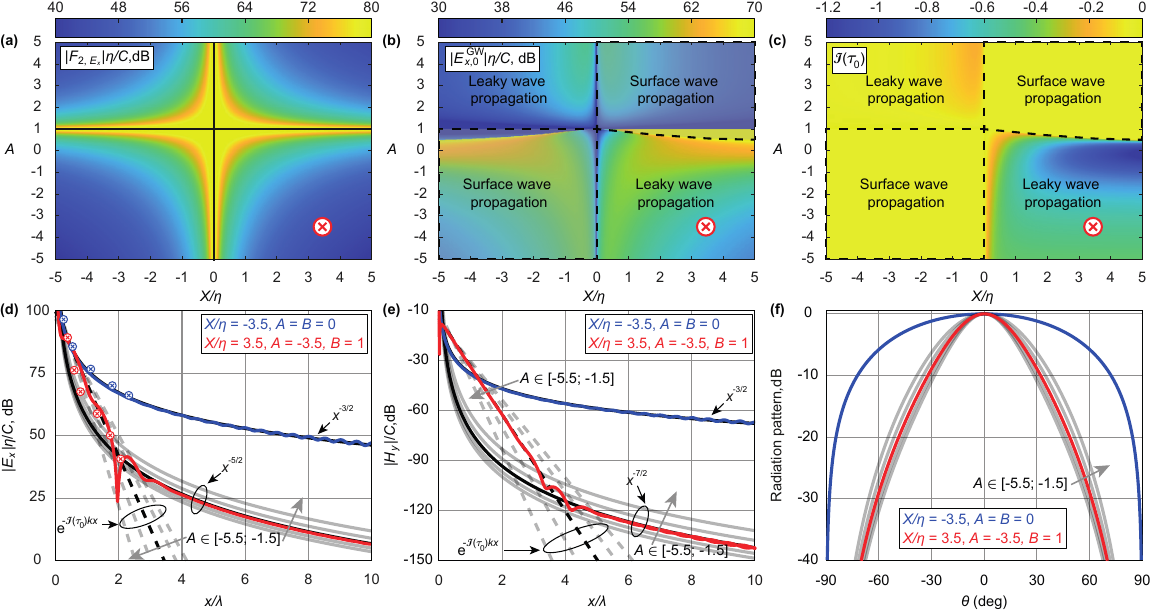} 
  \caption{Analytical study of fields excited by a 2D HMD on a nonlocal MS with a surface impedance of (\ref{eq:aprox}) with $B=1$ (Norton wave suppressed) depending on $A$ and $X/\eta$: (a) normalized excitation coefficient level $|F_{2,E_x}|\eta/C$ of the term decaying as $E_x\sim\Omega^{-5/2}$; (b) normalized excitation coefficient level $|E_{x,0}^{\text{GW}}|\eta/C$ of a SW/LW; (c) exponential decay factor $\Im(\tau_0)$ of the LW (SW propagation regions are uniformly yellow); normalized levels of the tangential electric (d) and magnetic (e) field (solid lines) calculated with (\ref{eq:ift}) vs. distance from the source $x/\lambda$ compared with their dominating asymptotic expansion terms from (\ref{eq:total_ff}); (f) radiation pattern shape. The case with the set of parameters $B=1$, $A=-3.5$, $X/\eta=3.5$, chosen for the practical implementation, is indicated with cross markers in (a--c) and with red lines in (d--f). The local high-impedance case is shown in (d--f) with blue curves. Markers in (d) show the results of full-wave numerical confirmation for the practical MS implementation.}
  \label{Fig2:all_theory}
\end{figure*}

Following \cite{rahmeier2022zero,dugan2024surface,Zhuravlev2025}, we assume the MS to be lossless and represent its purely imaginary impedance $\ZS(\gamma)$ as a
rational function, for which a second-order Padé approximation with one zero and one pole is \cite{Zhuravlev2025}:
\begin{equation}
    Z_\text{s}(\gamma)=jX\frac{1-\sum_{n=1}^{\infty} A_n\gamma^{2n}}{1-\sum_{m=1}^{\infty} B_m\gamma^{2m}}\approx jX\frac{1-A\gamma^2}{1-B\gamma^2},
    \label{eq:aprox}
\end{equation}
where $A\ne 1$, $X\ne0$ (to keep the 
GO term zero) and $B\ne A$ (to keep the nontrivial SD law) are real coefficients of the second-order impedance boundary condition. Note that with $A=B=0$ the value of $\ZS$ becomes constant with $\gamma$ (hereafter referred to as a \textit{local} case). As seen from (\ref{eq:total_ff}), the electric and magnetic field amplitudes of the Norton wave $F_{1,\lbrace E_x,H_y\rbrace}$ (the term with $n=1$) are  proportional to the second-order derivatives of $f_{E_x}$ and $f_{H_y}$, respectively, at $\theta=90^{\circ}$. Taking into account (\ref{eq:aprox}) the derivatives can be expressed as:
\begin{equation}
f^{''}_{H_y}=-\frac{2\eta^2}{X^2}\frac{(B-1)^2}{(A-1)^2};~ 
f^{''}_{E_x}=\frac{2j\eta^2}{X}\frac{(B-1)}{(A-1)}.
    \label{eq:2derivative}
\end{equation}
Both expressions (\ref{eq:2derivative}) show that the Norton wave cancels with the choice $B=1$ (a pole of the impedance is set at the grazing angle $\theta=90^{\circ}$). In this case, the next nonzero contributions in (\ref{eq:total_ff}) are $E_x\sim\Omega^{-5/2}$ and $H_y\sim\Omega^{-7/2}$, as well as of a single guided wave (with a complex propagation factor of $\tau_0$ satisfying the characteristic equation). As noted in \cite{Zhuravlev2025}, $ F_{0,H_y} $ determines the radiation pattern, which for $B=1$ is proportional to $\cos^3\theta$, as follows from \eqref{eq:ift_func} and \eqref{eq:aprox}. For comparison, the radiation pattern for the PMC is proportional only to $\cos\theta$. This means that the Norton wave suppression is expected to be accompanied by the radiation pattern narrowing. The corresponding excitation coefficients $F_{2,E_x}$, $F_{3,H_y}$, $E_{x,0}^{\text{GW}}$, and $H_{y,0}^{\text{GW}}$ as well as propagation factor $\tau_0$ can be analytically calculated using the formulas from the Supplemental Material \cite[Eq. (S.1)-(S.5)]{suppmat}.

The analytically calculated levels of $F_{2,E_x}$ (for the contribution $E_x\sim\Omega^{-5/2}$), as well as the excitation coefficient $E_{x,0}^{\text{GW}}$ and the decay factor $\Im(\tau_0)$ of the GW, are shown in Fig.~\ref{Fig2:all_theory}(a--c) in the case $B=1$ (Norton wave suppressed) for different $A$ and $X/\eta$. As follows from Fig.~\ref{Fig2:all_theory}(a), by simultaneously increasing both $|A|$ and $|X|$, one can decrease the contribution $E_x\sim\Omega^{-5/2}$ to the overall tangential electric field profile. On the other hand, for some combinations of $A$ and $X$, a propagating surface wave may appear (see the yellow regions in Fig.~\ref{Fig2:all_theory}(c) that must be avoided). For other combinations of $A$ and $X$, the guided wave becomes an LW with $\Im(\tau_0)$ strongly dependent on both $A$ and $X$. Although for $X<0$ and $A>0$ the LW has a relatively small amplitude, it has the slowest decay. Therefore, to accelerate the overall decay along the MS, it is preferable to select larger positive $X$ together with negative $A$. 

The contributions of the asymptotic term with  $E_x\sim\Omega^{-5/2}$ (solid gray curves) and LW (dashed gray curves) for different negative values of $A$ are plotted versus distance $x/\lambda$ along the MS in Fig.~\ref{Fig2:all_theory}(d). For the particular case $A=-3.5$, the asymptotic profiles are shown with black curves, while the total tangential electric field profile calculated using (\ref{eq:ift}) is shown with the solid red curve. It is clearly seen that at small distances ($x/\lambda<1.5$) the field decay is mainly determined by the LW, while at large distances ($x/\lambda>3$) it strictly follows the $E_x\sim\Omega^{-5/2}$ profile. In the middle range of distances, an interference of different wave contributions can be observed. A similar conclusion on the dominance of the term $H_y\sim\Omega^{-7/2}$ or LW depending on the distance can be drawn for the magnetic field profile shown in Fig.~\ref{Fig2:all_theory}(e). For comparison, the profiles of electric and magnetic fields excited by the same source calculated using (\ref{eq:ift}) are presented on the same plots for the local high-impedance MS with $X/\eta=-3.5$ along with asymptotic terms \cite[Eq. (S.6)-(S.7)]{suppmat}. Note that in this case, guided waves cannot be excited by the TM-polarized source due to capacitive impedance \cite[Par. 6.4.5]{TretyakovAMAE}, hence the field profiles precisely follow the contribution of the Norton wave ($E_x\sim\Omega^{-3/2}$). This comparison shows that for $x/\lambda>2$, despite some LW excitation, the field levels on the nonlocal MS with $B=1$ remain drastically lower than for the local MS, which can be explained by the Norton wave suppression. Furthermore, Fig.~\ref{Fig2:all_theory}(f) confirms that the Norton wave suppression on nonlocal MSs is associated with a significantly narrower beam (especially for larger $|A|$) pointed in the normal direction.

\begin{figure*}
  \centering
  \includegraphics[width=1\textwidth]{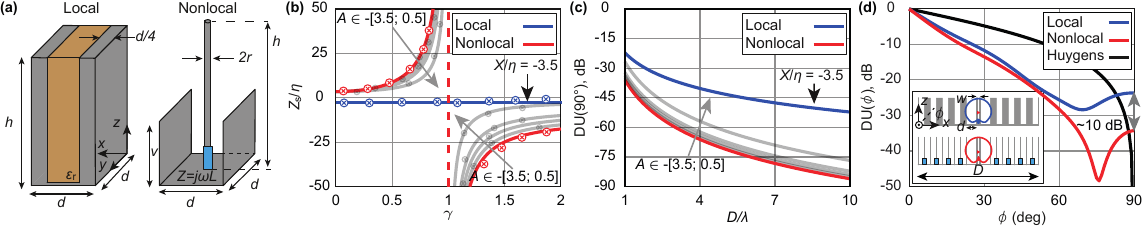} 
  \caption{Comparison of practical implementations of the local and nonlocal MSs: (a) meta-atoms of the local corrugated HIS (left) and the proposed nonlocal MS (right); (b) numerically extracted (markers) and analytically calculated (solid lines) $\ZS(\gamma)$; (c) analytically estimated $\text{DU}(90^{\circ})$ for $X/\eta=3.5$, $B=1$, and different practically implementable negative values of $A$ (see \cite[Tab. S.1]{suppmat}); (d) full-wave numerically calculated $\text{DU}(\phi)$ for $D=\lambda$.}
  \label{Fig3:CST_results}
\end{figure*}

To confirm the analytical predictions of the field decay acceleration, we numerically compare the excitation of two practical periodic structures implementing local and nonlocal MSs with the same period $d=\lambda/10=3$~mm at a frequency of $f=10$~GHz and the same thickness $h=6.2$~mm.
The local MS is implemented as a corrugated surface \cite{kildal1990artificially,Scire-Scappuzzo2009} with dielectric-filled ($\varepsilon_\text{r}=1.6$) PEC grooves (see Fig.~\ref{Fig3:CST_results}(a)).
The nonlocal MS is implemented as a modification of the HIS formed by vertical metal pins of height $h$ \cite{Silveirinha2008,Maslovski2010,Kaipa2011}, where the pins with a diameter of $2r=0.1$~mm are connected to the ground plane through lumped loads $Z=j\omega L$ with inductance $L$, and vertical PEC walls with a height of $v \le h$ and a thickness of 18~$\mu$m are placed between the adjacent pins. 

$\ZS$ as a function of $\gamma$ is parametrically calculated in CST Studio Suite for both types of meta-atoms using  periodic boundary conditions. The numerical results are shown in Fig.~\ref{Fig3:CST_results}(b) with markers being in good agreement with the analytical results obtained with (\ref{eq:aprox}). At 10~GHz (above the quarter-wave resonance), the local MS (see the blue curve and markers) exhibits a high and $\gamma$-independent capacitive impedance $X/\eta\approx-3.5$ \cite[Par. 11.4]{Collin1990}.
In contrast, for the nonlocal MS, the family of gray curves corresponding to $-3.5\le A \le-0.5$ for fixed $B=1$, $X/\eta=3.5$ exhibits a common pole at $\gamma=1$. This individual variation of $A$ can be achieved in the proposed structure by adjusting $v$ and $L$ simultaneously. Details of the impedance extraction method and the correspondence obtained between $v,L$ and $A$ are given in the Supplemental Material \cite[Sec. 2]{suppmat}. In particular, the case with $A\approx-3.5$ corresponds to $v=h=6.2$~mm and $L=0.63$~nH (see the red curve in Fig.~\ref{Fig3:CST_results}(b)). 

To further verify the analytically predicted field decay acceleration, we compared the numerically calculated electric field profiles of the latter nonlocal MS (with $A=-3.5$) with the local one. With this aim, in CST Studio Suite (full-wave simulation with Frequency Domain Solver), we introduced a 2D HMD  implemented as an internally fed open-ended parallel-plate waveguide with thickness $w=0.18$~mm embedded in the center of the structure with finite size $D$ (diameter along the $x$ axis). This finite 1D row containing $N$ meta-atoms is assumed to be periodically repeated in the $y$ direction. For both MSs, the open end is located in the top plane of the structure. To ensure the validity of the averaged boundary conditions, several closest meta-atoms at distances smaller than one period $d$ from the source are removed, as schematically depicted in the inset of Fig.~\ref{Fig3:CST_results}(d). The full-wave numerical results (markers) for $D=12\lambda$ and $N=118$ are compared with the analytical calculations (solid red and blue curves) in Fig.~\ref{Fig2:all_theory}(a) showing good agreement. Note that at distances from the source $x/\lambda>2$, the electric field decays by more than 70~dB, so that the expected decay profile $E_x\sim\Omega^{-5/2}$ cannot be confirmed for the practical structure due to the insufficient accuracy of numerical calculations (mainly caused by the effects of radiation boundaries). However, even at smaller distances the advantage in field decay along the nonlocal MS of more than 25~dB is clearly reached compared to the local case.

One possible application of the observed effect is mitigating edge diffraction, which typically limits the shielding action of finite-sized antenna reflectors. To evaluate the role of SD, let us compare the same two structures as above, but for different diameters $D$.
The shielding action can be quantified based on the far-field pattern using the so-called down-to-up ratio: $\text{DU}(\phi)=H_y^{\text{ff}}(-\phi)/H_y^{\text{ff}}(\phi)$, where $H_y^{\text{ff}}$ denotes the magnetic field in the far-field region; $\phi=\pi/2-\theta$. As shown in the Supplemental Material \cite[Sec. 3]{suppmat}, $\text{DU}(90^{\circ})$ can be analytically estimated using the physical optics approach. While the magnetic field in the zenith ($\theta=0^{\circ}$) is determined by $f_{H_y}(0^{\circ})$, the field in the nadir ($\theta=180^{\circ}$) can be approximated by the sum of two edge diffraction waves. The amplitude of both waves is proportional to the integral of the magnetic surface current density that would be induced by the HMD at an infinite MS (over the range of distances $D/2\le |x|<\infty$, where the finite structure of diameter $D$ is absent). For the nonlocal MS with $X/\eta \gg1$ and $B=1$ (the Norton wave suppressed), one can write:
\begin{equation}
    \label{eq:DU_nonlocal}
    \text{DU}^{\text{NL}}(90^{\circ})~\text{[dB]}\approx16.61-20\lg(1-A)-50\lg kD.
\end{equation}
A similar expression for the local HIS (with negative $\Im(\ZS)$ and $|X|/\eta \gg1$), for which the Norton wave dominates at large distances, reads \cite{Tatarnikov2005}:
\begin{equation}
    \label{eq:DU_local}
    \text{DU}^{\text{L}}(90^{\circ})~\text{[dB]}\approx1.05-30\lg kD.
\end{equation}

The analytical estimates for the local and nonlocal MSs (for different negative $A$) are plotted in Fig.~\ref{Fig3:CST_results}(c) versus $D$. As can be seen, the DU ratio in the nonlocal case with $A=-3.5$ can be reduced to -35~dB for a diameter of only one wavelength, which is by $\approx14$~dB better than for the local HIS of the same size. Full-wave numerical simulations of local and nonlocal practical structures with $D=\lambda$ ($N=8$) result in the $\text{DU}(\phi)$ curves shown in Fig.~\ref{Fig3:CST_results}(d). An improvement of 10~dB can be observed at $\phi=90^{\circ}$. A 4~dB deviation from the analytical estimation is due to the accuracy of the physical optics approximation and due to neglecting the contribution of the LW in the analytical estimation of the DU. Compared with a cardioid-shaped radiation pattern of a classical Huygens' source, both the local and nonlocal compact reflectors provide a steeper radiation pattern's roll-off near the horizon ($\phi=0^{\circ}$), which is a benefit in another important metric of  shielding action.

The results obtained clearly demonstrate the possibility of accelerating field decay along nonlocal MSs with second-order SD. Once the surface impedance has a pole at grazing angles, the contribution of the Norton wave in the asymptotic near-surface field expansion is suppressed. Although the contribution of the LW remains dominant near the HMD, the next nonzero higher-order term with profiles of $E_x\sim\Omega^{-5/2}$ and $H_y\sim\Omega^{-7/2}$ remains dominant at large distances from the source. This decay acceleration in comparison to a local HIS (with the Norton wave prescribing the decay profile of $E_x\sim\Omega^{-3/2}$) is possible to achieve with the proposed practical periodic structure of inductively-loaded pins separated with metal walls, which is found to exhibit the required specific law of SD. Excitation of the proposed nonlocal MS as a wavelength-size reflector is associated with edge diffraction reduced by 10~dB compared to a state-of-the-art local MS reflector of the same size. The investigated properties of nonlocal MSs, therefore, pave the way for efficient miniaturized shields and isolating structures that overcome the properties of conventional widely used corrugated surfaces.

\textit{Acknowledgment} --- This work was supported by the Russian Science Foundation (Project No.
25-19-00712, https://rscf.ru/en/project/25-19-00712/).

\bibliography{apssamp}
 \clearpage
 


\section*{Supplementary material}

# Supplemental Material: Accelerating field decay along nonlocal metasurfaces by suppressing the Norton wave

Alexander Zhuravlev,[1, *] Dmitry Tatarnikov,[1] Yury Kurenkov,[1] and Stanislav Glybovski[1]

[1]*School of Physics and Engineering, ITMO University, St. Petersburg, Russia*

## I. TERMS FROM THE FAR-FIELD APPROXIMATION OF THE FIELD

In this section, we explicitly present the power terms contained in Eq. (3) from the main article when the observation angle ($\theta$) tends to 90°. In this case, assuming that $(z-h)^2 \ll x^2$ and $h \ll \lambda$, the distance from the source $r$ can be approximated as $x$.

In the case when the Norton wave is suppressed, the power terms determining the field decay along the boundary are:

$$E_x^{-5/2}(x) = \underbrace{C\frac{j\eta^2}{X^3}\left[\frac{(1-A)X^2(A(B+2)-4B+1)+3(1-B)^3\eta^2}{(1-A)^3}\right]}_{F_{2,E_x}} \frac{e^{-j(k|x|-\frac{\pi}{4})}}{(k|x|)^{5/2}} \tag{S.1}$$

$$H_y^{-7/2}(x) = \underbrace{C\frac{j\eta^2}{X^6}\left[\frac{K_0+K_1+K_2}{(1-A)^6}\right]}_{F_{3,H_y}} \frac{e^{-j(k|x|-\frac{\pi}{4})}}{3(k|x|)^{5/2}} \tag{S.2}$$

where $C = I_0^{\mathrm{m}} k / \sqrt{2\pi}\eta$, $I_0^{\mathrm{m}}$ is the complex amplitude of magnetic line current, $k$ is the wavenumber of free space, $\eta$ is the characteristic impedance of free space, and coefficients $K_i$ are given by:

$$K_0 = (1-A)^2 X^4 (A^2(2B^2-34B+77)+2A(13B^2-86B+28)+107B^2-64B+2)$$
$$K_1 = 30(1-A)(1-B)^3 X^2 \eta^2 (A(B+5)-7B+1)$$
$$K_2 = 45(1-B)^6 \eta^4$$

The impact of guided waves (GW) is:

$$E_{x,0}^{\mathrm{GW}}/(2\pi jC) = \frac{-j\eta X(1-A\sin^2\tau_0)\cot\tau_0}{2jXA\cos\tau_0 + 3\eta B\cos^2\tau_0 + \eta(1-B)} \tag{S.3}$$

$$H_{y,0}^{\mathrm{GW}}/(2\pi jC) = \frac{\eta(1-B\sin^2\tau_0)\cot\tau_0}{2jXA\cos\tau_0 + 3\eta B\cos^2\tau_0 + \eta(1-B)} \tag{S.4}$$

The propagation constant $\tau_0$ of the GW can be calculated from the transverse resonance condition ($\eta\cos\tau + Z_\mathrm{s}(\tau) = 0$) [1], which can be written in the following form, using the fact that $\cos^2\tau + \sin^2\tau = 1$:

$$\cos^3\tau + \frac{jXA}{\eta B}\cos^2\tau + \frac{(1-B)}{B}\cos\tau + \frac{jX(1-A)}{\eta B} = 0 \tag{S.5}$$

Depending on the roots $\tau_i$ of the equation above, guided waves may be surface waves (SWs for $\sin\tau_i \in \mathbb{R}$) or leaky waves (LWs for $\sin\tau_i \in \mathbb{C}$). The equation above usually has three roots: one root corresponds to a SW and the others correspond to a LW decaying/rising along $x$ direction (see Fig. S.1). The brown circles in the figure highlight the positions of the root $\tau_0$ that is crossed when the original contour of integration P is deformed to the steepest–descent contour $\mathrm{P_{SD}}$; this GW can propagate along the boundary [2, Ch. 4].

* a.zhuravlev@metalab.ifmo.ru

In the case of local artificial magnetic conductor boundary ($|X/\eta| \gg 1$ and $A = B = 0$ in Eq. (4) from the main article) the first non-zero terms of Eq. (3) from the main article are:

$$E_x^{-3/2}(x) = \underbrace{-C\frac{\eta^2(1-B)}{X(1-A)}}_{F_{1,E_x}} \frac{e^{-j(k|x|-\frac{\pi}{4})}}{(k|x|)^{3/2}} \tag{S.6}$$

$$H_y^{-3/2}(x) = \underbrace{-C\frac{j\eta^2}{X^2}\frac{(1-B)^2}{(1-A)^2}}_{F_{1,H_y}} \frac{e^{-j(k|x|-\frac{\pi}{4})}}{(k|x|)^{3/2}} \tag{S.7}$$

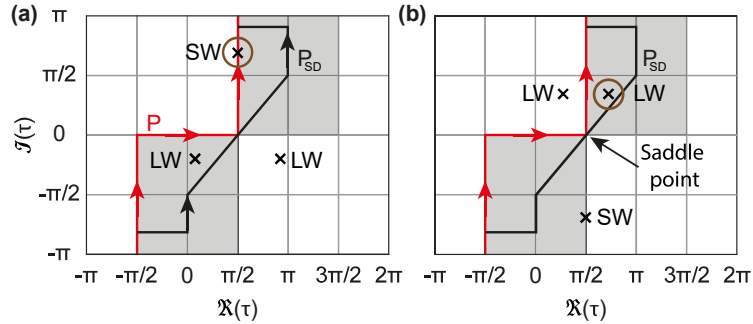

FIG. S.1. Typical location of $\tau_i$ on the complex plane $\tau$: (a) corresponds to the excitation of a SW and (b) corresponds to the excitation of a LW. The brown circles in the figure highlight the positions of the root $\tau_0$ that is crossed when the original contour of integration P is deformed to the steepest–descent contour $P_{SD}$; this GW can propagate along the boundary.

## II. PERIODIC STRUCTURE OF INDUCTIVELY-LOADED PINS SEPARATED WITH METAL WALLS

To demonstrate how the height of the vertical dividers affects the spatial dispersion of $Z_s(\gamma)$, where $\gamma = k_x/k = \sin\theta$, associated with the structure shown in Fig. S.2(a), we calculated its $Z_s(\gamma)$ for different values of $v$ keeping the Norton wave suppression condition satisfied ($B = 1$). The impedance under normal incidence was chosen to be $3.5\eta$ which demanded $h = 6.22$ mm. The following parameters were locked: $2r = 0.1$ mm, $d = 3$ mm, while the value of inductance $L$ was adjusted to control the angular response.

To obtain spatially dispersive surface impedance, we calculated the angular dependence of the reflection coefficient $\rho(\gamma)$, from the unit cell using CST Studio Suite (CST) with periodic boundary conditions (see Fig. S.2(b)). Spatially dispersive surface impedance was obtained via formula:

$$Z_s(\gamma) = \eta\sqrt{1-\gamma^2}\frac{1+\rho(\gamma)}{1-\rho(\gamma)} \tag{S.8}$$

The normalized wave vector tangential component of the incident plane wave reads: $\gamma = \sin\theta + 2\pi n/kD$ [3, par. 4.1], where $D$ is the size of the calculation domain along the $x$–axis and $n$ is a mode number. As the structure is self-periodic, we can say that $D = dN$, where $d$ is unit cell periodicity and $N$ is a number of unit cells. Considering fundamental mode ($n = 0$), one can see that $\gamma < 1$. On the other hand, the normalized tangential component of the wave vector, associated with the first high order mode ($n = 1$), can be arbitrarily controlled by the adjusting of $N$ and $\theta$. The last fact was used to estimate surface impedance for evanescent waves. At the last step the obtained $Z_s(\gamma)$ curve was fitted by Eq. (4) from the main article to get $X$, $A$, and $B$ coefficients.

The results of parametric study are shown in the Tab. S.1. One can see that the increase of $v$ can decrease the value of $A$ and implementable values of $A \in [-3.5; -0.5]$ for considered geometrical parameters.

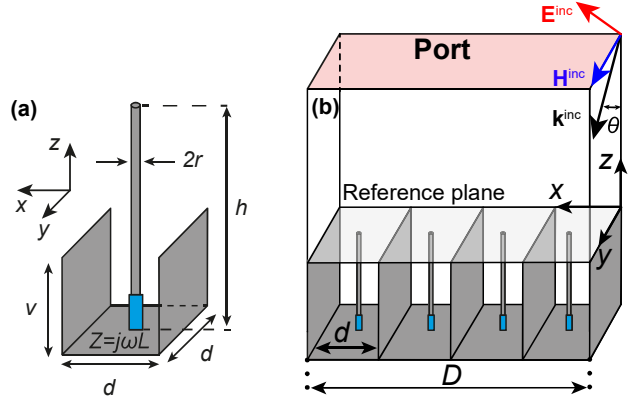

FIG. S.2. (a) Periodic structure of inductively-loaded pins separated with metal walls. (b) Simulation setup for surface impedance extraction in CST.

TABLE S.1. The dependence of $X$, $A$, and $B$ coefficients on the structure's parameters.

| $v$, mm | $L$, nH | $X/\eta$ | $A$ | $B$ |
|---|---|---|---|---|
| 6.22 | 0.63 | 3.5 | -3.5 | 1 |
| 5.92 | 0.68 | 3.5 | -3.0 | 1 |
| 5.70 | 0.74 | 3.5 | -2.5 | 1 |
| 5.44 | 0.81 | 3.5 | -1.9 | 1 |
| 3.12 | 1.68 | 3.5 | -0.5 | 1 |

## III. DOWN-TO-UP RATIO ESTIMATION WITH THE PHYSICAL OPTICS APPROACH

Here, we present an approximate method of calculation down-to-up ratio ($\mathrm{DU}^{\mathrm{NL}}(\phi=90°) = H_y^{\mathrm{ff}}(-90°)/H_y^{\mathrm{ff}}(90°)$, where $H_y^{\mathrm{ff}}$ denotes the magnetic field in the far-field region and $\phi = \pi/2 - \theta$) of the system shown in Fig. S.3. The distant magnetic field of the source shielded by a finite screen of length $D$ can be approximated as $H_y^{\mathrm{ff}} = H_y^{\infty} - H_y^{\mathrm{cut}}$, where the first term is the radiation of the infinite shielded antenna and the second one is the radiation of the cut parts. The radiation of the cut parts can be found using physical optics as:

$$H_y^{\mathrm{cut}} = \int_{-\infty}^{-L/2} \left( H_y^{\mathrm{e}} + H_y^{\mathrm{m}} \right) dx + \int_{L/2}^{\infty} \left( H_y^{\mathrm{e}} + H_y^{\mathrm{m}} \right) dx \tag{S.9}$$

where we assume that each element $\Delta x$ radiates as a combination of elementary 2-D sources [2, Par. 5.4]:

$$H_y^e = -J_0^e \frac{k\sqrt{2}}{4\sqrt{\pi}} \frac{e^{-j(\Omega - \frac{\pi}{4})}}{\sqrt{\Omega}} \cos\theta\, e^{jkx\sin\theta} \Delta x \tag{S.10}$$

$$H_y^m = -J_0^m e \frac{k\sqrt{2}}{4\eta\sqrt{\pi}} \frac{e^{-j(\Omega - \frac{\pi}{4})}}{\sqrt{\Omega}} e^{jkx\sin\theta} \Delta x$$

where $J_0^e$ or $J_0^m$ is electric or magnetic surface current densities (A/m or V/m), $\Omega = kr$, and $r = \sqrt{x^2 + z^2}$.

If we assume that cutting the shield does not affect the field distributions along it, the currents which radiate can be found through the fields distributions along the shield through the boundary conditions: $\mathbf{J}^{\mathrm{m}} = -\mathbf{z}_0 \times \mathbf{E}$ and $\mathbf{J}^{\mathrm{e}} = \mathbf{z}_0 \times \mathbf{H}$.

As shown above, a specific case when $B = 1$ can provide such field decay along the shield:

$$E_x \approx -\frac{C}{e^{-j\frac{\pi}{4}}} \frac{j3\eta^2}{X(1-A)} \frac{e^{-jk|x|}}{(k|x|)^{5/2}} - E_{x,0}^{\mathrm{GW}} e^{-jk|x|\sin\tau_0} \tag{S.11}$$

$$H_y \approx \frac{j15C}{e^{-j\frac{\pi}{4}}} \frac{\eta^2}{X^2(1-A)^2} \frac{e^{-jk|x|}}{(k|x|)^{7/2}} - H_{y,0}^{\mathrm{GW}} e^{-jk|x|\sin\tau_0}$$

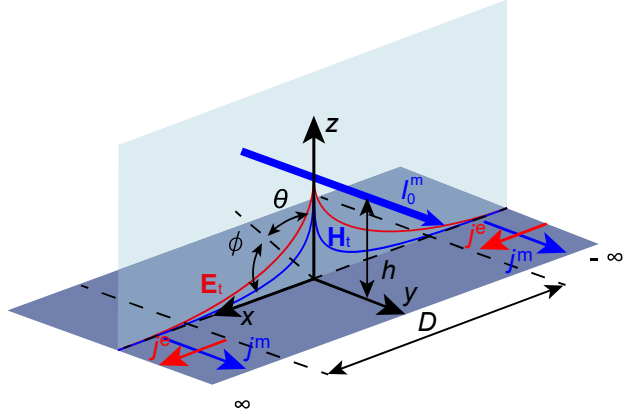

FIG. S.3. Infinite screen is cut from both sides at $x = \pm D/2$. The cut parts radiate as elementary 2-D sources.

Because the magnetic field decays faster than the electric one along the screen, we can neglect the radiation of the electric currents in (S.9). Moreover, to achieve a simple equation, we presume that the contribution to (S.9) is mainly due to the power term $E_x^{-5/2} \sim x^{-5/2}$ in (S.11) and we can neglect the effect of GW.

Using the boundary condition for electric field and then putting magnetic current in (S.10), one can write (S.9) as:

$$H_y^{\text{cut}} = C' \left( \int_{-\infty}^{-kL/2} \frac{e^{-jk|x|+jkx\sin\theta}}{(k|x|)^{5/2}} dx + \int_{kL/2}^{\infty} \frac{e^{-jk|x|+jkx\sin\theta}}{(k|x|)^{5/2}} dx \right) \quad (S.12)$$

where $C' = -C \frac{j3\eta^2}{X(1-A)} e^{j\frac{\pi}{4}} \frac{k\sqrt{2}}{4\eta\sqrt{\pi}} \frac{e^{-j(\Omega-\frac{\pi}{4})}}{\sqrt{\Omega}}$.

Considering the radiation in normal direction ($\theta = 0°$), we can neglect the radiation of the cut parts $H_y^{\text{ff}}(\theta = 0°) \approx H_y^{\infty}(\theta = 0°)$ (see Eq. (3) from the main article):

$$H_y^{\text{ff}}(\theta = 0°) \approx -C \frac{\eta}{jX + \eta} \frac{e^{-j(\Omega-\frac{\pi}{4})}}{\sqrt{\Omega}} \quad (S.13)$$

The radiation directly below the shield is determined by the cut parts $H_y^{\text{ff}}(\theta = 180°) \approx -H_y^{\text{cut}}(\theta = 180°)$. Putting the observation angle $\theta = 180°$ in (S.12) and making substitution $s = kx$, we arrive to:

$$H_y^{\text{ff}}(\theta = 180°) \approx -\frac{C'}{k} 2 \int_{kL/2}^{\infty} \frac{e^{-js}}{s^{5/2}} ds \approx \frac{jC'}{k} 2 \frac{e^{-jkL/2}}{(kL/2)^{5/2}} \quad (S.14)$$

After dividing the last equation by (S.13), $\text{DU}^{\text{NL}}(90°)$ can be found as:

$$\text{DU}^{\text{NL}}(90°) = j\sqrt{\frac{9}{2\pi}} \frac{jX+\eta}{X(1-A)} \frac{e^{j\pi/4} e^{-jkL/2}}{(0.5kL)^{5/2}} \quad (S.15)$$


\end{document}